\documentclass[showpacs,preprintnumbers,aps]{revtex4}
\begin{document}
\preprint{USM-TH-169}
\title {A note on the static potential for a D=3 Born-Infeld theory
coupled to a new generalized connection}
\author{Patricio Gaete}
\email {patricio.gaete@usm.cl}
\affiliation{Departamento de F\'{\i}sica, Universidad T\'ecnica F.
Santa Mar\'{\i}a, Valpara\'{\i}so, Chile\\ and \\
Departamento de F\'{\i}sica, Universidad Tecnol\'ogica
Metropolitana , Santiago, Chile}
\date{\today}
\begin{abstract}
For a (2+1)-dimensional Born-Infeld theory coupled to a recently
proposed generalized connection, we compute the interaction
potential within the structure of the gauge-invariant but
path-dependent variables formalism. The result is equivalent to
that of $QED_3$ with a Thirring interaction term among fermions,
in the short distance regime. This result agrees with that of the
topologically massive Born-Infeld theory.
\end{abstract}
\pacs{ 11.10.Ef, 11.15.-q}
\maketitle

\section{Introduction}

In recent years, the nonlinear Born-Infeld electrodynamics have
generated a lot of interest due to their appearance in $D$-branes
physics. It is a well known fact that, for instance, the low
energy dynamics of $D$-branes have been described by a nonlinear
Born- Infeld type action \cite{Tseytlin, Gibbons}. It is worth
recalling here that Born and Infeld \cite{Born} suggested to
modify Maxwell electrodynamics to get rid of infinities of the
theory. In addition to the string interest, the Born-Infeld theory
has also attracted attention from different viewpoints. For
example, in connection to noncommutative field theories
\cite{Gomis}, also in magnetic monopoles studies \cite{Kim}, and
possible experimental determination of the parameter that measures
the nonlinearity of the theory \cite{Denisov}.

On the other hand, we further recall that systems in $(2+1)$
dimensions have been extensively discussed in the last few years
\cite{Deser, Dunne, Khare}. This is primarily due to the
possibility of realizing fractional statistics, where the physical
excitations obeying it are called anyons, which continuously
interpolate between bosons and fermions. In this respect, the
three-dimensional Chern-Simons gauge theory is the key example, so
that Wilczek's charge-flux composite model of anyon can be
implemented \cite{Wilczek}. Interestingly, we call attention to
the fact that recently a novel way to describe anyons has been
discussed \cite{Itzhaki}. In this case, the crucial
ingredient for this is the introduction of a generalized
connection which permits to realize fractional statistics, such
that the Chern-Simons term needs not be introduced. Indeed, by
using a Lagrangian which describes Maxwell theory coupled to the
current via the generalized connection leads to fractional
statistics by the same mechanism, as in the case of the
Maxwell-Chern-Simons theory, of attaching a magnetic flux to the
electrons \cite{Itzhaki, Gaete1}.

We also recall that in recent times a great of deal of attention
has been devoted to the concept of duality by its unifying role in
physics. As is well known, duality refers to an equivalence
between two or more quantum field theories whose corresponding
classical theories are different. Consequently, one obtains more
information about a theory than is possible by considering a
single description. The archetype example on this equivalence is
the well known duality between the self-dual and
Maxwell-Chern-Simons theories in $(2+1)$ dimensions
\cite{Townsend, Jackiw}, which is obtained via the master
Lagrangian approach to show the dynamical equivalence between
both theories. Other interesting approaches have been used in
different theories, such as the nonlinear $BF$ models and
higher-order derivative models, to establish the duality mapping
 \cite{Clovis}. It is, however, desirable to have some additional
check on duality from a somewhat different perspective. We are
primarily concerned with the physical content associated to
duality. In fact, one can verify the existence of duality between
two apparently different theories by computing physical quantities
in both theories. If the two answers agree the theories should be
dual. This approach is the main focus of this note.

With these ideas in mind, by computing the interaction potential
between trial external charges, we have showed the equivalence
between a $(2+1)$-dimensional topologically massive Born-Infeld
theory \cite{Tripathy, Harikumar} and the Maxwell-Chern-Simons
theory with a Thirring interaction term among fermions, in the
short distance regime \cite{Gaete2}. As already mentioned, the
aim of the present paper is to further explore this point. With
this in view, we propose to extend the previous analysis to a
$(2+1)$-dimensional Born-Infeld theory coupled to a generalized
connection, which permits fractional statistics. Again, as in the
case of the topologically massive Born-Infeld theory, we find that
the static potential for this new theory, too, agrees with that of the
Maxwell-Chern-Simons theory with a Thirring interaction term
among fermions. One is thus lead to the interesting conclusion
that the lowest-order modification of the static potential due to the
presence of the Born-Infeld term, simulates the effect of
including the self-interaction term in the Maxwell-Chern-Simons
theory. This is our main result. In this way we establish a new
equivalence between both theories, in the hope that this will be
helpful to gain a better understanding of effective theories in lower
dimensions.

As already stated, our objective is to compute explicitly the
interaction energy between static pointlike sources for a
$(2+1)$-dimensional gauge theory containing both the Born-
Infeld and the new generalized connection terms. The starting
point is the Lagrangian:
\begin{equation}
{\cal L} = \beta ^2 \left[ {1 - \sqrt {1 + \frac{1}{{2\beta ^2
}}F_{\mu \nu } F^{\mu \nu } } } \right] + \frac{\theta
}{2}\varepsilon _{\mu \nu \rho } J^\mu  F^{\nu \rho }  - A_\mu
J^\mu, \label{bigc10}
\end{equation}
where $J^\mu$ is the external current and $\theta$ is a parameter
with dimension $M^{-1}$. The parameter $ \beta $
measures the nonlinearity of the theory and in the limit $ \beta
\to \infty $ the Lagrangian (\ref{bigc10}) reduces to the Maxwell
plus a generalized connection term. In order to handle the square
root in the Lagrangian (\ref{bigc10}), we introduce an auxiliary
field $v$, such that its equation of motion gives back the
original theory\cite{Tseytlin}. This allows us to rewrite the
Lagrangian (\ref{bigc10}) in the form
\begin{equation}
{\cal L} = \beta ^2 \left[ {1 - \frac{1}{2}v\left( {1 +
\frac{1}{{2\beta ^2 }}F_{\mu \nu } F^{\mu \nu } } \right) -
\frac{1}{{2v}}} \right] + \frac{\theta }{2}\varepsilon _{\mu \nu
\rho } J^\mu F^{\nu \rho }  - A_\mu  J^\mu. \label{bigc20}
\end{equation}
Once this is done, the canonical quantization of this theory from
the Hamiltonian point of view is straightforward and follows
closely that of Refs. \cite{Gaete1, Gaete2}.
The canonical momenta read $\Pi ^\mu = vF^{\mu 0} +
\theta\varepsilon ^{0\mu \nu } J_\nu$, which result in the usual
primary constraint $\Pi ^0=0$ and $ p \equiv \frac{{\partial {\cal
L}}}{{\partial {\dot v}}} = 0$. The canonical Hamiltonian
following from the above Lagrangian
\begin{equation}
\begin{array}{r}
 H_C  = \int {d^2 } x\left\{ { - \frac{1}{{2v}}\left( {\Pi ^i \Pi _i
- 2\theta \varepsilon ^{ij} \Pi _i J_j  + 2\theta ^2 J_j J^j  -
\beta ^2 } \right) + \frac{v}{2}\left( {\frac{1}{2}F_{ij} F^{ij}
+ \beta ^2 } \right)} \right\} +  \\ + \int {d^2 } x\left\{ { -
A_0 \left( {\partial _i \Pi ^i  - J^0 } \right) - \frac{\theta
}{2}\varepsilon _{ik} F^{ik} J^0  + A_k J^k - \beta ^2} \right\}.
\\ \label{bigc30}
\end{array}
\end{equation}

Requiring the primary constraint $\Pi ^0=0$ to be stationary, we
obtain the secondary constraint
$\Gamma_1(x)\equiv\partial _i \Pi ^i - J^0=0$. The consistency
condition for the $p$ constraint yields no further constraints and
just determines the field $v$,
\begin{equation}
v = \sqrt {\frac{{1 - \frac{1}{{\beta ^2 }}\left( {\Pi ^i \Pi _i -
2\theta \varepsilon ^{ij} \Pi_i J _j + 2 {\theta ^2 }J^j J_j }
\right)}}{{1 + \frac{1}{{2\beta ^2 }}F^{ij} F_{ij} }}},
\label{bigc40}
\end{equation}
which will be used to eliminate $v$. The extended Hamiltonian that
generates translations in time then reads $H = H_C  + \int d x
\left( {c_0 (x)\Pi_0 (x) + c_1 (x)\Gamma _1 (x)} \right)$, where
$c_0(x)$ and $c_1(x)$ are the Lagrange multipliers. Since $ \Pi_0
= 0$ for all time and $ \dot{A}_0 \left( x \right) = \left[ {A_0
\left( x \right),H} \right] = c_0 \left( x \right)$, which is
completely arbitrary, we discard $ A_0 \left( x \right)$ and $ \Pi
_0 \left( x \right)$ because they add nothing to the description
of the system. As a result, the Hamiltonian becomes
\begin{equation}
H  = \int {d^2 } x\left\{ {\beta ^2 \sqrt {\left( {1 + \frac{{B^2
}} {{\beta ^2 }}} \right)\left( {1 + \frac {{{\bf D}^2 }}{{\beta
^2 }}} \right)}  - \beta ^2 -\theta B  J^{0} +A_{k}J^{k} - c^
{\prime }  \left(
 x \right)\left( {\partial _i \Pi ^i- J^0 } \right)} \right\},
\label {bigc50}
\end{equation}
where $c^ \prime  \left( x \right) = c_1 \left( x \right) - A_0
\left( x \right)$, $B^2  = \frac{1}{2} F_{ij}F^{ij}$ and ${\bf
D}^2 = - \left( {\Pi _i \Pi ^i  + 2\theta \varepsilon ^{ik} J_i
\Pi _k + 2 \theta ^2 J_i J^i } \right)$.

Since there is one first class constraint $\Gamma_1(x)$ (Gauss' law),
we choose one gauge fixing condition that will make the full set of
constraints becomes second class. We choose the gauge fixing
condition to correspond to
\begin{equation}
\Omega _2 \left( x \right) \equiv \int\limits_{C_{\xi x} } {dz^\nu
} A_\nu  \left( z \right) = \int_0^1 {d\lambda } x^i A_i \left(
{\lambda x} \right) = 0, \label{bigc60}
\end{equation}
where  $\lambda$  $\left( {0 \le \lambda  \le 1} \right)$ is the
parameter describing the spacelike straight path between the
reference points $ \xi ^k $ and $ x^k $ , on a fixed time slice.
For simplicity we have assumed the reference point $\xi^k=0$. The
choice (\ref{bigc60}) leads to the Poincar\'{e} gauge
\cite{Gaete2}. With this we obtain the only nontrivial Dirac bracket
\begin{equation}
\left\{ A_{i}(x),\pi ^{j}(y)\right\} ^{*}=g _{i}^{j}\delta ^{(2)}
\left( x-y\right) -\partial _{i}^{x}\int_{0}^ {1}d\lambda
x^{j}\delta ^{(2)}\left( \lambda x-y\right) .  \label{bigc70}
\end{equation}

Since we are interested in estimating the lowest-order correction
to the interaction energy, we will retain only the leading
quadratic term in the expression (\ref{bigc50}). Thus the
Hamiltonian may be written as
\begin{equation}
H = \int {d^2 } x\left\{ {\frac{1}{2}\left( {{\bf D}^2  + B^2 }
\right)\left[ {1 - \frac{1}{{4\beta ^2 }}\left( {{\bf D}^2  + B^2
} \right)} \right] + \frac{1}{{2\beta ^2 }}B^2 {\bf D}^2  - \theta
BJ^0  + A_k J^k } \right\}. \label{bigc80}
\end{equation}
Notice that $\bf D$ refers to the field $ D^i\equiv vE^i = \Pi ^i
- \theta \varepsilon ^{ij} J_j$. With the help of the Dirac
bracket (\ref{bigc70}), we can now write the Dirac brackets in
terms of $ B=\varepsilon _{ij}\partial ^{i}A^{j}$ and $ D^i  = \Pi
^i  - \theta \varepsilon ^{ij} J_j$ fields, that is,
\begin{equation}
\left\{ {D_i \left(  x \right),D_j \left(  y \right)} \right\}^ * = 0 ,
\label{bigc90a}
\end{equation}
\begin{equation}
\left\{ {B \left(  x \right),B \left(  y \right)} \right\}^ * = 0 ,
\label{bigc90b}
\end{equation}
\begin{equation}
\left\{ {D_i \left( x \right),B\left(  y \right)} \right\}^ * = -
\varepsilon _{ij} \partial _x^j \delta ^{(2)} \left
( x - y \right). \label{bigc90c}
\end{equation}
It must now be observed that, unlike the
Maxwell-Chern-Simons-Born- Infeld theory \cite{Gaete2}, in the
present model , the bracket (\ref{bigc90a}) is commutative. It gives
rise to the following equations of motion for $D_i$ and $B$ fields:
\begin{equation}
{\dot D}_i \left( x \right) = \int {d^2 } y\left( {1 - \frac{1}{
{2\beta ^2 }}\left( {B^2  - {\bf D}^2 } \right)B - \theta J^0 }
\right) \left\{ {D_i \left( x \right),B\left( y \right)} \right\}^
* + J_i \left( x \right) + \int {d^2 } yJ^k \left( y \right)\partial _k
\int_0^1 {d\lambda } y_i \delta ^{\left( 2 \right)} \left( {\lambda
 x - y} \right), \label{bigc100}
\end{equation}
\begin{equation}
{\dot B}\left(  x \right) = \int {d^2 } y\left( {1 - \frac{1}{{2
\beta ^2 }}\left( {{\bf D}^2  - B^2 } \right)} \right)\left\{
{B\left( x \right),\frac{1}{2}{\bf D}^2 \left( y \right)}
\right\}^ *. \label{bigc110}
\end{equation}
In the same way, we write the Gauss law as:
\begin{equation}
\partial _i D_L^i  = J^0, \label{bigc120}
\end{equation}
where $D_L^i$ refers to the longitudinal part of $D^i$. For $J^{i}=0$,
the static electromagnetic fields assume the form
\begin{equation}
vB(x) - \theta J^0  = 0, \label{bigc130}
\end{equation}
\begin{equation}
\varepsilon _{ij} \partial _i D_j \left( x \right) = 0. \label{bigc140}
\end{equation}
Note that Gauss' law for the present theory reads $\partial_{i}D^{i}=
J^{0}$, in other words,
\begin{equation}
D^i  = \partial ^i \left( { - \frac{{J^0 }}{{\nabla ^2 }}} \right), \label{bigc150}
\end{equation}
where $\nabla ^2$ is the two-dimensional Laplacian. It is
worthwhile recalling at this point that the field $v$ at leading
order in $\beta$ reads
\begin{equation}
v = 1 + \frac{1}{{2\beta ^2 }}\left( {{\bf D}^2  - B^2 } \right).
\label{bigc160}
\end{equation}
Then using $D^i=vE^i$, and for $ J^0
\left( {t,{\bf x}} \right) = e\delta ^ {(2)} \left( {\bf x} \right)$, the electric field
at leading order in $\beta$  takes the form
\begin{equation}
E^i  = \left\{ {1 - \frac{1}{{2\beta ^2 }}\left( {\frac{e}{{2\pi }}\frac{{x_i }}{
{|{\bf x}|^2 }}} \right)^2 } \right\}\partial ^i \left( { - \frac{{J^0 }}{
{\nabla ^2 }}} \right), \label{bigc170}
\end{equation}
where, as in the case of the topologically massive Born-Infeld theory,
we have dropped a divergent factor in expression (\ref{bigc170}) which
does not contribute in the limit $\beta\gg\theta$.

Now we are prepared to compute the potential energy for a pair
of static pointlike opposite charges at ${\bf y}$ and ${\bf
y}^\prime$ along the lines of Refs. \cite{Gaete1, Gaete2}. In such
a case, we start by considering
\begin{equation}
V \equiv e\left( {{\cal A}_0 \left( \bf y \right) - {\cal A}_0
\left( {\bf y^ \prime} \right)} \right), \label{bigc180}
\end{equation}
where the physical scalar potential ${\cal A}_0$ is expressed in
terms of the electric field
\begin{equation}
{\cal A}_0 \left( {t,\bf x} \right) = \int_0^1 {d\lambda } x^i E_i
\left( {t,\lambda \bf x} \right).
\label{bigc190}
\end{equation}
The following remark deserves to be mentioned: Eq.(\ref{bigc190})
follows from the vector gauge- invariant field \cite{Gaete3}:
\begin{equation}
{\cal A}_\mu  \left( x \right) \equiv A_\mu  \left( x \right) +
\partial _\mu \left( { - \int_\xi ^x
{dz^\mu  } A_\mu  \left( z \right)} \right), \label{bigc200}
\end{equation}
where, as in Eq.(\ref{bigc60}), the line integral appearing in the
above expression is along a spacelike path from the point $\xi$ to
$x$, on a fixed time slice. At the same time, it should be noted
that the gauge-invariant variables (\ref{bigc200}) commute with the
sole first class constraint (Gauss' law), supporting the fact that
these fields are physical variables \cite{Dirac}.

Using Eq.(\ref{bigc170}), we rewrite Eq.(\ref{bigc190}) as
\begin{equation}
{\cal A}_0 \left( {t,{\bf x}} \right) = \int_0^1 {d\lambda }
\left\{ {1 - \frac{1}{2}\left( {\frac{e}{{2\pi \beta }}} \right)^2
\left( {\frac{{\lambda x_i }}{{|\lambda {\bf x}|^{2}}}} \right)^2
} \right\} x^i \partial _i^{\lambda {\bf x}} \left( { - \frac{{J^0
\left( {\lambda {\bf x}} \right)}}{{\nabla _{\lambda {\bf x}}^2
}}} \right) \label{bigc210}
\end{equation}
where $J^0$ is the external current. The static current describing
two opposite charges $e$ and $-e$ located at $\bf y$ and $\bf y
\prime$ is then given by $J^0 \left( {t,{\bf x}} \right) =
e\left\{ {\delta ^{\left( 2 \right)} \left( {{\bf x} - {\bf y}}
\right) - \delta ^{\left( 2 \right)} \left( {{\bf x} - {\bf y}^
\prime  } \right)} \right\}$. Substituting this back into Eq.
(\ref{bigc210}) we get accordingly
\begin{equation}
V = \frac{{e^2 }}{\pi }\ln \left( {\mu |{\bf y} - {\bf y}^
{\prime} |} \right) - \frac{{e^4 }}{{ 16\pi ^3 \beta ^2
}}\frac{1}{{|{\bf y} - {\bf y}^ {\prime} |^2 }}, \label{bigc220}
\end{equation}
where $\mu$ is a cutoff with mass dimension introduced to regularize
the potential.

As mentioned earlier, this is exactly the result obtained for
$QED_3$ with a Thirring interaction term among fermions, the
so-called generalized Maxwell-Chern-Simons gauge theory
\cite{Gaete3,Ghosh}, in the short distance regime. This fact has
been noted previously for the topologically massive Born-Infeld
theory \cite{Gaete2}. Incidentally, it is of interest to notice
that the result based on the generalized Maxwell-Chern-Simons
theory has been derived from the bosonized version of a $U(1)$
gauged massive Thirring model \cite{Ghosh}. This allows us to
interpret the model proposed here as a effective theory which
contains quantum effects at the classical level. One thus obtains
a similarity between the tree level mechanism exploited here and
the bosonization tool used in Ref. \cite{Ghosh}. This observation
and the result (\ref{bigc220}) are new.

We briefly summarize the results so far obtained. Our present
model, though very simple, has been found to produce results which
strongly simulate the general characteristics of the $QED_{3}$
with a Thirring interaction among fermions, exactly as it happens
with the topologically massive Born-Infeld theory. It is quite
appealing that there are a class of models which can predict the
interaction energy (\ref{bigc220}). Finally, the methodology
presented here provides a physically-based alternative to the
usual Wilson loop approach.

\section{ACKNOWLEDGMENTS}

Work supported in part by Fondecyt (Chile) grant 1050546.

\end{document}